\documentclass[]{emulateapj}
\usepackage{epsfig, natbib, graphicx, color}

\input epsf

\newcommand{\Mpc}{\mbox{Mpc}}

\newcommand{\msun}{M_\odot}

\newcommand{\avgn}{\langle \bar n \rangle}

\newcommand{\nm}{ \frac{d\avgn}{dm}}

\newcommand{\Nt}{N_{T}}

\newcommand{\Nobs}{N_{obs}}

\newcommand{\Nsat}{N_{sat}}

\newcommand{\avg}[1]{\left\langle #1 \right\rangle}

\newcommand{\lk}{{\cal{L}}}

\shortauthors{ROZO ET AL.}
\shorttitle{COSMOLOGICAL CONSTRAINTS FROM CLUSTERS}

\begin{document}
\title{Cosmological Constraints from SDSS maxBCG Cluster Abundances}
\author{Eduardo Rozo \altaffilmark{1,2,3}, 
Risa H. Wechsler\altaffilmark{3,4}, 
Benjamin P. Koester\altaffilmark{5,6}, 
Timothy A. McKay\altaffilmark{6,7}, 
August E. Evrard\altaffilmark{6,7}, 
David Johnston\altaffilmark{8}, 
Erin S. Sheldon\altaffilmark{9},
James Annis\altaffilmark{10},
Joshua A. Frieman\altaffilmark{3,5,10}
}

\altaffiltext{1}{CCAPP, The Ohio State University, Columbus, OH 43210, erozo@mps.ohio-state.edu}
\altaffiltext{2}{Department of Physics, The University of Chicago, Chicago, IL 60637}
\altaffiltext{3}{
  Kavli Institute for Cosmological Physics,
  The University of Chicago, 
  Chicago, IL 60637 USA
}

\altaffiltext{4}{
  Kavli Institute for Particle Astrophysics \& Cosmology,
  Physics Department, and Stanford Linear Accelerator Center,
  Stanford University,
  Stanford, CA 94305
}
\altaffiltext{5}{Department of Astronomy, The University of Chicago, Chicago, IL 60637}
\altaffiltext{6}{Department of Physics, University of Michigan, Ann
  Arbor, MI 48109}
\altaffiltext{7}{Astronomy Department, University of Michigan, Ann
  Arbor, MI 48109}
\altaffiltext{8}{Jet Propulsion Laboratory, Caltech, Pasadena, CA
  91109}
\altaffiltext{9}{Center for Cosmology and Particle Physics and
  Department of Physics, New York University, New York, NY 10003}
\altaffiltext{10}{Center for Particle Astrophysics, Fermilab, Batavia, IL 60501}
\begin{abstract}
We perform a maximum likelihood analysis of the cluster abundance
measured in the SDSS using the maxBCG cluster finding algorithm.
Our analysis is aimed at constraining the power spectrum
normalization $\sigma_8$, and assumes flat cosmologies with a scale
invariant spectrum, massless neutrinos, and CMB and supernova priors
$\Omega_m h^2=0.128\pm0.01$ and $h=0.72\pm0.05$ respectively.
Following the method described in the companion paper 
\citet[][]{rozoetal07a}, we derive
$\sigma_8=0.92\pm0.10$ ($1\sigma$) after marginalizing over all
major systematic uncertainties.  We place strong lower limits on the
normalization, $\sigma_8>0.76$ ($95\%$ CL) ($>0.68$ at $99\%$ CL).
We also find that our analysis favors relatively low values for the
slope of the Halo Occupation Distribution (HOD),
$\alpha=0.83\pm0.06$.  The uncertainties of these determinations
will substantially improve upon completion of an ongoing campaign to
estimate dynamical, weak lensing, and X-ray cluster masses in the
SDSS maxBCG cluster sample.
\end{abstract}


\section{Introduction}

One of the most important problems in observational cosmology today is
to resolve the question of whether or not dark energy takes the form
of a cosmological constant.  While current geometric
probes of dark energy such as supernovae and baryon acoustic
oscillations unequivocally tell us that dark energy exists, a
complementary probe of the dark energy evolution can help us
distinguish between a cosmological constant and a dynamical dark
energy.  The growth of structure is one such probe
\citep[][]{ekeetal96,holderetal01,evrardetal02, molnaretal04}.  In
particular, given a geometric probe, an accurate determination of the
amplitude of the power spectrum at two different times directly
constrains the growth between the two epochs, and can in principle
help determine if the dark energy density evolves with redshift.  
We know the power spectrum amplitude at the time of last
scattering with high precision thanks to Cosmic Microwave Background
(CMB) experiments \citep[see e.g.][and references therein]{wmap06}, so
a precise determination of the current power spectrum normalization may,
in principle, distinguish between a dynamical dark energy component
and a cosmological constant.  

It is well known that the abundance of massive halos in the local
universe depends strongly on the amplitude of the matter power
spectrum $\sigma_8$.\footnote{Here, we characterize the present day
amplitude of the power spectrum with the usual parameter $\sigma_8$,
the rms amplitude of density perturbations in spheres of $8h^{-1}\
\Mpc$ radii.}  In particular, from theoretical considerations
\citep[][]{pressschechter74,bondetal91,shethtormen02} one expects the
number of massive clusters in the universe to be exponentially
sensitive to $\sigma_8$, a picture that has been confirmed with
extensive numerical simulations.  Consequently, the number of galaxy
clusters within a given survey region ought to be able to provide
powerful constraints on $\sigma_8$, and, indeed, the literature is
rife with these type of studies.

Unfortunately, cluster abundance determinations of $\sigma_8$ have to
overcome a large variety of difficulties.  For instance, the abundance
of massive halos in the universe is sensitive not only to $\sigma_8$
but also to $\Omega_m$, the mean matter density of the universe, which
implies that constraints on the number density of massive halos
typically result in large degeneracies between $\Omega_m$ and
$\sigma_8$ \citep[though see][]{rozoetal04}.  In fact, the main
the main obstacle to accurate $\sigma_8$ measurements is systematic
uncertainties in mass estimates of clusters \citep[see
e.g.][]{pierpaolietal03,henry04}.  Consequently, new analyses that
properly marginalize over such systematic uncertainties are of
particular importance to interpret cluster abundance constraints
within a broad cosmological context.

In this work, we use the techniques developed in
\citet[][]{rozoetal07a} to analyze the Sloan Digital Sky Survey (SDSS)
maxBCG cluster sample from \citet[][]{koesteretal06a}.  The analysis
of these data uses information from the \citet[][]{rozoetal07a}
companion paper on the form of the maxBCG selection function, which
connects our observable mass proxy --- the cluster richness --- to
halo mass. As discussed in in that work, at this time our
understanding of the maxBCG selection function is incomplete, which
implies that we have large systematic uncertainties in the selection
function.  Nevertheless, the large mass range probed by the maxBCG
cluster sample allows us to marginalize over these uncertainties and
still recover competitive estimates for $\sigma_8$, albeit with the
inclusion of cosmological priors for $\Omega_mh^2$ and $h$ in our
analysis.  Importantly, the results presented in this work ought to be
interpreted as an upper limit of how well we can expect $\sigma_8$ to
be constrained from the current maxBCG cluster sample in the near
future.  Not only will we soon be able to include additional data such
as weak lensing and dynamical cluster mass estimates, but we also
expect our understanding of the maxBCG cluster selection function to
improve both as an expanded suite of simulations used to calibrate the
maxBCG selection function become an even more accurate approximation
to reality and as we work towards a more robust richness estimator.

The layout of the paper is as follows.  In \S \ref{sec:algorithm}
we briefly describe the maxBCG cluster finding algorithm used to
identify the cluster sample analyzed in this work.  \S
\ref{sec:model} summarizes the model described in
\citet[][]{rozoetal07a}, including a description of the various
parameters and priors used in our analysis.  Our cosmological constraints
are presented in \S \ref{sec:results}, and discussed in detail in \S
\ref{sec:discussion}.


\section{The maxBCG Cluster-Finding Algorithm}
\label{sec:algorithm}

Details of how the maxBCG cluster-finding algorithm works can be found
in \citet[][]{koesteretal06a}.  Here, we only summarize the main
elements of the cluster-finding algorithm.

MaxBCG is an optical cluster-finding algorithm that relies on
photometric measurements to overcome projection effects.  To detect
clusters, maxBCG uses the well known observational fact that galaxy
clusters contain a large number of so called \it ridgeline \rm
galaxies: bright, red, early type galaxies that populate a narrow
ridgeline in color-magnitude space as a function of redshift.  The
color distribution of these galaxies is modeled as a narrow Gaussian,
while their two dimensional spatial distribution about the cluster
center is modeled as a projected Navarro, Frenk, and White (NFW)
profile \citep[][]{NFW}.  To determine the cluster center, maxBCG
relies on the observational fact that, in the vast majority of
clusters, there is a clear Brightest Cluster Galaxy (BCG) at or near
the center of the cluster.  These BCG galaxies tend to be extremely
luminous and red, populating the tip of the color-magnitude ridgeline.
Their luminosity and colors are also modeled as narrow Gaussians.

Given our model, one can then compute the likelihood that a particular
galaxy is the BCG galaxy of a cluster by computing the product
$\lk_{BCG}\lk_{R}$, where $\lk_{BCG}(z)$ is the likelihood that the
galaxy under consideration has the observed colors and magnitude
assuming that it is a BCG at redshift $z$, and $\lk_{R}$ is the
likelihood that the galaxy distribution around the the candidate BCG
will occur under the assumption that the BCG is at the center of
cluster (though this is unlikely to be universally true; see
e.g. \citealt{vandenbosch05}).  In computing the ridgeline likelihood
$\lk_{R}$, only galaxies found within $3\ \Mpc$ of the candidate BCG
and in a narrow ($3\sigma$) color window around the expected color for
ridgeline galaxies at that redshift are considered. The total
likelihood $\lk=\lk_{BCG}\lk_{R}$ is evaluated at a grid of redshifts,
and a photometric redshift estimate for the cluster is obtained by
maximizing the likelihood function.

The end result of this process is a likelihood assignment for every
galaxy in the survey.  The galaxy list is then rank ordered according
to likelihood, and the most likely BCG is selected as a BCG.  A first
rough richness measurement is made by counting the number of galaxies
within $1\ \Mpc$, which is then used to estimate a characteristic
radius for the cluster using the results from
\citet[][]{hansenetal05}.  All ridgeline galaxies above a luminosity
cut of $0.4L_*$ within this scale radius are considered cluster
members, and the number of cluster members is defined as the cluster's
richness, denoted here by $\Nobs$ and in \citet[][]{koesteretal06a} by
$N_{gals}^{r_{200}}$ \footnote{For technical reasons, in this work the
  luminosity cut for $\Nt$ is defined in the $r$-band, whereas $\Nobs$
  has an $i$-band luminosity cut.}.  All BCG candidates within a
galaxy overdensity of 200 and within $\Delta z=0.02$ of the cluster just
found are dropped from the candidate BCG list.  The procedure is then
iterated, resulting in a cluster catalog where each cluster has an
assigned cluster center, a members' list, a richness, and a
photometric redshift estimate.


\section{The Model}
\label{sec:model}

In this work, our main observable is the number of clusters within the
survey volume as a function of cluster richness, i.e. the richness function.
We now summarize the basic picture behind our analysis.  A detailed presentation
of this formalism can be found in \citet[][]{rozoetal07a}.

\subsection{The Model at a Glance}

\begin{figure}[t]
\epsscale{1.2}
\plotone{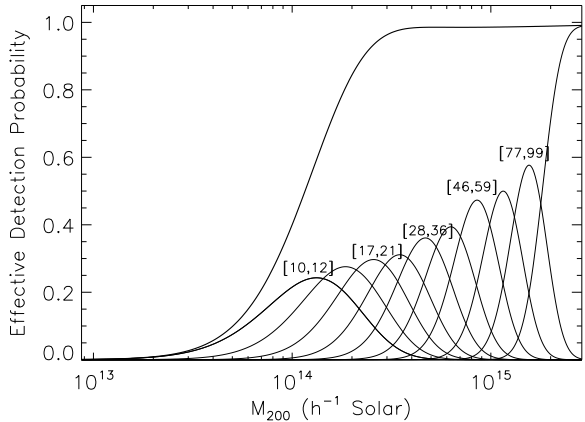}
\caption[]{Effective mass selection function for the maxBCG SDSS cluster sample
of \citet[][]{koesteretal06a} in our maximum likelihood model.  The thin solid lines
are the selection function for each of our 10 richness bins.  The 
selection function for the richest bin continues to grow at larger masses due to 
the extrapolation of the decreasing purity function\citep[see][]{rozoetal07a}.   
Note, however, that all abundances remain finite since the halo 
mass function is decreasing exponentially fast at these mass scales.  The thick,
solid line is the net sum of the individual bins, and gives the effective mass selection
for the cluster sample as a whole.}
\label{fig:mass_binning}
\end{figure}

Suppose we wish an expression for the number of clusters of a given
richness.  In general, if $P(\Nobs|m)$ is the probability that a halo
of mass $m$ is detected as a cluster with $\Nobs$ galaxies, the
mean density of these clusters is simply
\begin{equation}
\avg{n} = \int dm\ \nm P(\Nobs|m).
\end{equation}
The main idea behind our analysis is to split $P(\Nobs|m)$ into two:
there is a probability $P(\Nt|m)$, namely the Halo Occupation Distribution or HOD, 
that determines the intrinsic scatter between a halo's mass and its richness,
and a second distribution $P(\Nobs|\Nt)$ that characterizes the observational
scatter.  If $c(\Nt)$ is the probability that a halo with $\Nt$ galaxies
is detected, then the probability that a halo of mass $m$ is detected as a cluster
with $\Nobs$ galaxies is
\begin{equation}
P(\Nobs|m) = \sum_{\Nt} c(\Nt)P(\Nobs|\Nt)P(\Nt|m).
\end{equation}
Note that, by definition, $c(\Nt)$ is also the expected fraction of detected halos, 
and hence we refer to it as the completeness.
Finally, suppose $p(\Nobs)$ is the probability of a cluster with $\Nobs$ galaxies being a real detection, i.e. of the cluster corresponding to an actual halo.
By definition, $p(\Nobs)$ is also the expected fraction of clusters that are real, and 
hence we call $p(\Nobs)$ the purity of the
sample. If $p(\Nobs)\neq 1$, then the observed number
of clusters will be boosted relative to our above estimate by a factor 
$1/p(\Nobs)$, so the mean number density of the clusters becomes
\begin{equation}
\avg{n} = \int dm\ \nm \psi(m)
\end{equation}
where
\begin{equation}
\psi(m) = \frac{1}{p(\Nobs)}\sum_{\Nt} c(\Nt)P(\Nobs|\Nt)P(\Nt|m).
\end{equation}
The quantity $\psi(m)$ is the effective mass binning of our cluster
sample, that is, $\psi(m)$ is the fraction of mass $m$ clusters that fall into
the richness bin of interest.  Note, however, that when $p(\Nobs)<1$, then
$\psi(m)$ can be larger than unity.  The true mass binning is obtained by setting
$p(\Nobs)=1$.  Figure \ref{fig:mass_binning} shows the effective mass binning of
of our cluster sample assuming the maximum likelihood values of all relevant
model parameters as determined by our analysis (see below). 


\subsection{The Likelihood Function}
\label{sec:lkhd}

The above model for cluster abundances tells us what the expected
cluster abundance will be.  For our analysis, however, we wish to know
the likelihood of observing a particular data set given some
cosmological and HOD parameters.  As described in
\citet[][]{rozoetal07a}, we model the probability of observing a
realization given a set of cosmological and HOD parameters as a
Gaussian.  While more accurate likelihood functions can be found in
the literature \citep[][]{hucohn06,holder06}, these ignore
correlations due to scatter in the mass--observable relation, and thus
we have opted for a simple Gaussian model, which is expected to hold
if bins are sufficiently wide to include a large ($\gtrsim 10$) number
of clusters.

The contributions to the correlation matrix that we consider are:
\begin{enumerate}
\item A Poisson contribution due to the Poisson fluctuation in the
number of halos of mass $m$ within any given volume. 
\item A sample variance contribution reflecting
the fact that the survey volume may be slightly overdense or underdense with respect
to the universe at large.
\item The bin-to-bin scatter arising from the stochasticity of $\Nobs$ as a function of $m$.
\item A contribution due to the statistical uncertainties associated with photometric 
redshift estimation.
\item A contribution due to the stochastic nature of the completeness and purity functions. That
is, if we know the expected purity and completeness of an infinite sample, 
any finite sample may have slightly
different fractions of true detections and false positives.
\end{enumerate}

The detailed construction of this likelihood function can be found in
\citet[][]{rozoetal07a}.  For the purposes of this work, the most
important aspect of our likelihood function is that, as demonstrated
in \citet[][]{rozoetal07a}, our likelihood analysis correctly recovers
the underlying cosmological and halo occupation distribution
parameters to within the intrinsic degeneracies of the data for mock
maxBCG catalogs constructed using the techniques in
(Wechsler et al. 2007, in preparation).  Consequently, we are fully confident that
our analysis technique is sound, robust, and that it properly takes
into account the various systematic uncertainties that affect the
construction of the \citet[][]{koesteretal06a} maxBCG catalog.


\subsection{Model Parameters}
\label{sec:model_parameters}

For reference, we summarize here all of the relevant parameters for
our model.  The cosmological parameters we considered are $\sigma_8,
\Omega_m,$ and $h$.  The power spectrum is taken to be scale
invariant, and we use the low baryon transfer function from
\citet[][]{eisensteinhu99} with zero neutrino masses.  The baryon
density is fixed to the WMAP3 value in \citet[][]{wmap06}.  All results
reported in this work use a \citet{jenkinsetal01} mass function,
and we have explicitly checked that the expectation values for all parameters
of interest recovered using different mass function parameterizations 
\citep[in particular those of][]{shethtormen02,warrenetal05} fall well within
the $1\sigma$ error bars recovered using the \citet[][]{jenkinsetal01}
mass function.   We also
assume a flat $\Lambda CDM$ universe\footnote{Note that because of our 
cluster abundance determination of $\sigma_8$ is both local and uses
only a narrow redshift range, the constraints we recover from the
sample are largely independent of the dynamics of dark
energy. Consequently, we do not expect assuming a $\Lambda$CDM
universe will bias our results in any significant way.  As error
bars shrink, however, this assumption will undoubtedly need to be
relaxed.}.  Following \citet[][see also
\citealt{zhengetal05,yangetal05b}]{kravtsovetal04} we assume that the
total number of galaxies in a halo takes the form $\Nt=1+\Nsat$ where
$\Nsat$, the number of satellite galaxies in the cluster, is Poisson
distributed at each halo mass $m$ with an expectation value
$\avg{\Nsat|m}$ given by
\begin{equation}
\avg{\Nsat|m}= \left(\frac{m}{M_1}\right)^\alpha.
\label{eq:hod}
\end{equation}
Here, $M_1$ is the characteristic mass at which halos acquire one
satellite galaxy.  Note that in cluster abundance studies, the typical
mass scale probed is considerably larger than $M_1$.  Nevertheless,
the above parameterization is convenient because degeneracies between
HOD and cosmological parameters take on particularly simple forms when
parameterized in this way \citep[see][]{rozoetal04}.

Our model also includes a large number of nuisance parameters.  These
are described in detail in \citet[][]{rozoetal07a}.  There, we
demonstrate that the completeness function is flat as a function of
richness, and hence can be described with only one nuisance parameter
$c$.  The purity function, on the other hand, is clearly peaked, with
purity decreasing both in the high and low richness limits.  We found
that we can accurately describe the purity function with two
nuisance parameters, $p_0$ and $p_1$.  Two more parameters, $B_0$ and
$\beta$, describe the mean value of $\Nobs$ for clusters at fixed
$\Nt$.  These parameters characterize the amplitude and slope of the
mean relation $\avg{\Nobs|\Nt}$ respectively.  An additional parameter
$B_1$ describes the variance of $\Nobs$ at fixed $\Nt$, which is
taken to be simply proportional to $\avg{\Nobs|\Nt}$\footnote{An
example of such a distribution is Poisson statistics, in which case
the proportionality constant is simply unity.}.  The form of the
distribution $P(\Nobs|\Nt)$ is taken to be a discretized Gaussian
based on simulations (see below).  Finally, two additional nuisance
parameters $\avg{b_z}$ and $\sigma_{z}$ calibrate the bias and
scatter of our photometric redshift estimates.    A summary of the
equations that define all of our nuisance parameters is presented in
Appendix \ref{app:pardef}.


\subsection{Model Priors}
\label{sec:priors}

In \citet[][]{rozoetal07a}, we attempted to calibrate the various
nuisance parameters in our model using mock catalogs created with the
method of {Wechsler et al. 2007, in preparation).  We found that for some
parameters, systematic variation between simulations dominated over
random errors. Specifically, while both the completeness and purity
functions appeared to be stable and robustly constrained, the
parameters which characterizes the normalization of the probability
matrix $P(\Nobs|\Nt)$ and its scatter, $B_0$ and $B_1$ were seen to
have large systematic variations amongst the three mock catalogs
considered.  These systematic variations, however, appeared to fall
along a degeneracy band, as seen in the lower panel  of Figure 5 
of the companion paper \citet[][]{rozoetal07a}, and reproduced
here in Appendix \ref{app:pardef}.  Consequently, we
placed a weak prior on $B_0$ and $B_1$ along this degeneracy,
and wide enough that it comfortably encompassed the $95\%$
statistical regions of the parameters in all three simulations considered
in \citet[][]{rozoetal07a}.   Our prior is shown with solid lines in Figure 
\ref{fig:priors}.
The \it slope \rm $\beta$ of the mean relation
$\avg{\Nobs|\Nt}$ appeared to be somewhat better constrained at
roughly the $\approx 5\%-10\%$ level.  In an effort to be conservative, and
given the small number of simulations we had available, we have opted
for placing a more generous $15\%$ Gaussian prior on $\beta$ centered on
the simulation-calibrated value.  Since both
the completeness and purity functions appear to be robustly
constrained in the simulations, we also use the simulation-calibrated
priors from \citet[][]{rozoetal07a} for these quantities.  Finally, whereas
we found the range of photometric redshift parameters to also be
systematics dominated, it was clear that the range of these uncertainties
was small enough that photometric redshift uncertainties have a
minimal impact in our results.  Thus, we simply marginalize over the
range of photometric redshift parameters observed in the simulations.

Unfortunately, the above priors are simply not restrictive enough for
constraining cosmological parameters.  In particular, the range of
cosmological and HOD models is large enough that evaluation of the
likelihood function over the entire degeneracy region becomes
impossible.  To overcome this difficulty, we therefore include three
additional priors.  The first is a CMB based prior on the matter
density, $\Omega_m h^2=0.128\pm0.01$.  It is important to note,
however, that the CMB constraint on the matter density of the universe
is independent of the amplitude of the power spectrum, as it depends
only on the well known physics of the photon-baryon fluid of the early
universe.  Consequently, use of this prior in our cosmological
analysis should not introduce a bias in our estimate for $\sigma_8$ \it
regardless of the dynamical nature of dark energy. \rm In a similar
spirit, we assume a supernova-based prior $h=0.73\pm0.05$, and
finally, a generous theoretically-motivated prior on $\alpha$, the
slope of the HOD, which we take to be $\alpha=1.0\pm0.15$ \citep[see
e.g.][]{kravtsovetal04}.  All of these priors are assumed to be
Gaussian in log space (i.e. lognormal).

To summarize, then, we have placed priors on the physical parameters 
$\Omega_m h^2$, $h$, and $\alpha$, the nuisance parameter $\beta$, and 
the purity and completeness functions.  The amplitude and scatter of the
$\Nobs-\Nt$ relation are allowed to float essentially freely.  The remaining
physical parameters of interest are $M_1$, the mass scale of the HOD, and
$\sigma_8$, the amplitude of the power spectrum in cluster scales. 


\section{Results}
\label{sec:results}


\begin{figure}[t]
\epsscale{1.2}
\plotone{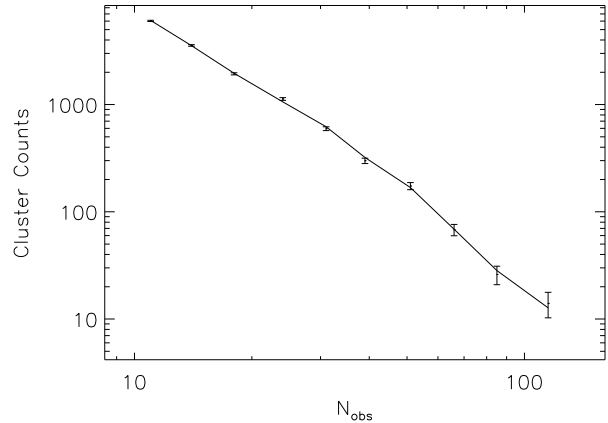}
\caption[Comparison Between maxBCG Cluster Counts and the
Maximum-Likelihood Model Fits]{Comparison between the observed binned 
cluster counts in the maxBCG catalog (solid circles) and our
maximum-likelihood model.   Error bars represent Poisson uncertainties, and
are shown for reference only as the counts in the various richness bins
are correlated.  The agreement between our maximum likelihood model and
the observed number counts is excellent.}
\label{fig:data_counts}
\end{figure} 


Constraints on our model parameters are obtained from the likelihood
function described above via a Monte Carlo Markov Chain (MCMC)
approach.  The details of our MCMC algorithm can be found in
\citet[][]{rozoetal07a}.  Briefly, leaning heavily on the work by
\citet[][]{dunkleyetal05}, we construct MCMCs that are optimal in step
size, and we ensure we make enough evaluations to robustly recover the
$95\%$ confidence likelihood contours in parameter space.  The data
itself is binned in nine logarithmic bins in the richness range
$100>\Nobs\geq 10$, plus an additional high richness bin containing
all clusters with $100$ galaxies or more.  The redshift range is
$[z_{min},z_{max}]=[0.1,0.3]$.  These bins are wide enough that even
our least-populated bin contains $14$ clusters, which is necessary
given our likelihood function\footnote{Specifically, the likelihood
approximates the Poisson uncertainty in the abundance as Gaussian,
so a large number of clusters per bin is necessary in our analysis}.
Finally, in order to make sure our results are robust, we ran two
additional MCMCs, and checked that the recovered distributions
were consistent with each other.  We found this to be the case.  
Consequently, we then proceeded to join all three chains into
a single chain with $3\cdot 10^5$ points.  This number of evaluations
is enough to recover the $95\%$ confidence regions of the distribution
with $\approx 5\%$ accuracy, or alternatively the $99\%$ confidence
regions with $\approx 10\%$ accuracy.   Finally, in order to test whether 
our results were robust to the number of clusters with most extreme 
richnesses, we also ran an MCMC where the 
richness range was limited to $100\geq\Nobs\geq 11$, which amounts to
dropping the 14 richest clusters and the 2558 clusters with $\Nobs=10$.
We found that this chain produced results consistent with those of our original
analysis.


\begin{figure}[t]
\epsscale{1.2}
\plotone{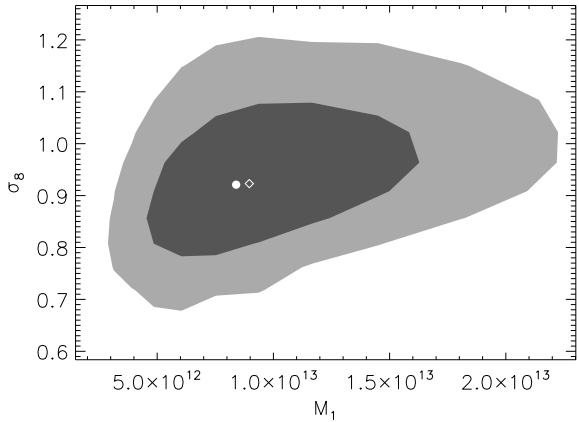}
\caption[Confidence Regions in the $\sigma_8-M_1$]{
Confidence regions ($68\%$ and 
$95\%$) in the $\sigma_8-M_1$
plane.  The circle marks the median values of $M_1$ and $\sigma_8$ 
$(8.4\cdot 10^{12} \msun,0.92)$, while
the diamond marks the corresponding average values $(9\cdot10^{12}\msun,0.92)$.
$\sigma_8$ and $M_1$ are the only two completely free parameters in our
analysis.}
\label{fig:data_s8_M1}
\end{figure} 


A comparison between the observed number counts and the recovered
maximum-likelihood model can be seen in Figure \ref{fig:data_counts}.
We can see that our best fit model provides an excellent fit to the
observed number counts.  The corresponding $68\%$ and $95\%$
confidence regions in the $\sigma_8-M_1$ plane are seen in Figure
\ref{fig:data_s8_M1}.  The median and average values for each of these
two parameters are $(M_1,\sigma_8)=(8.4\cdot10^{12}\msun,0.92)$ and
$(M_1,\sigma_8)=(9.0\cdot10^{12}\msun,0.92)$ respectively.
The marginalized distribution for $\sigma_8$, shown
in Figure \ref{fig:data_s8}, is roughly fit by a Gaussian distribution
with $\sigma_8=0.92\pm0.10$, and allows us to place an interesting
lower limit on $\sigma_8$: $\sigma_8 > 0.76$ ($95\%$ CL) or 
$\sigma_8>0.68$ ($99\%$ CL).  The $M_1$
distribution is roughly lognormal, though with some slight skewness.  The
corresponding $1-\sigma$ parameters are $\ln (M_1/10^{12} \msun) =
2.1\pm0.4$, corresponding to $M_1=8.2^{+4.0}_{-2.7}\ \times
10^{12}\msun$.  

While our remaining model parameters were all constrained with priors,
it is nevertheless interesting to look at their a posteriori
distributions.  In particular, we find that there appears to be some
tension between our prior on the slope $\beta$ of the $\Nobs-\Nt$
relation, and our prior on $\alpha$, the slope of the HOD.  This is
illustrated in Figure \ref{fig:data_aN_beta}, where we show the
posterior distribution in the $\alpha-\beta$ plane.  The circle marks 
the median values of the marginalized distributions 
$(\beta,\alpha)=(1.15,0.84)$
whereas the square marks the central values of our
priors, $\alpha=1.0$ and $\beta=1.18$.  The average values of the parameters
are very close to the median values.
The degeneracy direction is roughly
$\alpha\beta=constant$, as expected based on the fact that $\Nobs\sim
\Nt^\beta \sim m^{\alpha\beta}$.  While the width of our priors is
large enough that we do not feel this discrepancy biases our results,
it is clear from Figure \ref{fig:data_aN_beta} that either the slope
of the $\Nobs-\Nt$ relation is shallower than seen in the simulations,
or the slope of the HOD is markedly different from unity.  Information
from a new suite of simulations for currently ongoing work indicates that
it is probably the former: our prior for $\beta$ seems to be on the
high edge of what is possible. 


\begin{figure}[t]
\epsscale{1.2}
\plotone{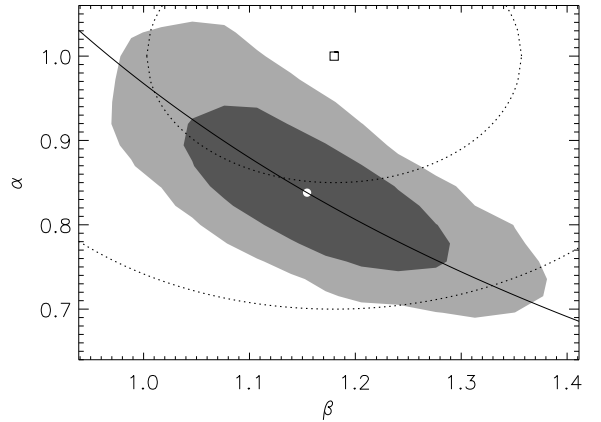}
\caption[]{The posterior distribution in the $\alpha-\beta$ plane.
  The filled contours represent the $68\%$ and $95\%$ confidence
  regions, which are centered on the square; 
  also shown for reference are the corresponding regions in
  our prior distribution.  The circle marks the median values of each
  of the parameters $(\beta,\alpha)=(1.15,0.84)$, which are almost
  identical to the average values.  
  Note that the central value for our input priors (the square) lies outside the
  $95\%$ confidence region of the a posteriori distribution, 
  suggesting that either our calibration is
  incorrect, or the slope of the HOD is substantially lower than
  unity.  The solid line corresponds to the expected
  $\alpha\beta\approx\ constant$ degeneracy between the two
  parameters.}
\label{fig:data_aN_beta}
\end{figure} 

 
While it is impossible for us to fully determine whether the HOD prior
or our simulation-calibrated prior is more incorrect without
additional information, we can try to better understand how each of
these two possibilities would affect our results.  To do so, we have
run two additional MCMCs, one with a tight, $5\%$ prior on $\alpha$
and no prior on $\beta$, and one with the converse priors.  Due to the
strong $\alpha\beta$ degeneracy, in either case we found that the
maximum likelihood model resulted in an excellent fit to the data.
 
Our results are summarized in Figure \ref{fig:data_s8}.  Briefly, we
find that the tight $\beta$ prior favors a low $\alpha$
($\alpha=0.76\pm0.05$) and a very high $\sigma_8$
($\sigma_8=1.05^{+0.13}_{-0.11}$), whereas the tight HOD prior
$\alpha=1.0\pm0.05$ favors a lower $\sigma_8$ value,
$\sigma_8=0.92\pm0.11$.  The implausibly large value for $\sigma_8$
and the further lowering of $\alpha$ for the $\beta$ prior 
suggests that our central value for $\beta$ may be too high, which is
consistent with more recent preliminary determinations of the maxBCG selection
function in simulations of our ongoing work.  


\begin{figure}[t]
\epsscale{1.2}
\plotone{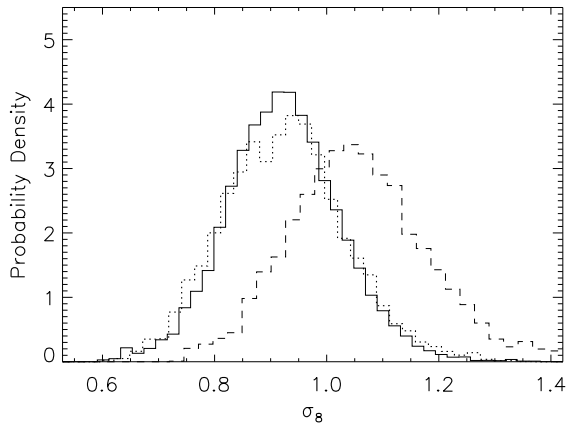}
\caption[]{The solid histogram shows the marginalized $\sigma_8$ distribution 
recovered from our MCMC.  A Gaussian fit to the distribution results in 
$\sigma_8=0.92\pm0.10$. From the above distribution we can also directly 
obtain a lower limit for
$\sigma_8$, $\sigma_8>0.76$ at $95\%$ confidence, or
$\sigma_8>0.68$ at $99\%$ confidence.  Also shown are the
marginalized distributions recovered when one places a tight, $5\%$ prior
on the slope of the HOD (dotted line) and no prior on the selection function,
and for the converse case (dashed line).}
\label{fig:data_s8}
\end{figure} 



\section{Discussion}
\label{sec:discussion}

\subsection{Comparison With Other Work}
\label{sec:comparison}

We have performed a detailed statistical analysis of the observed
cluster counts in the maxBCG cluster catalog of
\citet[][]{koesteretal06b}.  We find $\sigma_8=0.92\pm0.10$
(1-$\sigma$), and place an interesting lower limit $\sigma_8>0.75$
($95\%$ CL) or $\sigma_8>0.66$ ($99\%$ CL).  All of our results are
marginalized over all major systematic effects, though they are also
subject to additional cosmological priors $\Omega_m h^2=0.128\pm0.01$
and $h=0.73\pm0.05$.  Finally, our analysis was restricted to flat
cosmologies with massless neutrinos and scale-invariant primordial
power spectra.

Our value for $\sigma_8$ is somewhat larger --- but consistent with ---
the typical value obtained from cluster abundance studies using X-rays
\citep[$\sigma_8\approx 0.7$, see e.g.][and references
therein]{borganietal01,pierpaolietal01,ikebeetal02,allenetal03,pierpaolietal03,
  schueckeretal03,henry04}.  $\sigma_8$ determinations from optical
cluster samples are more rare, but here again we find our
results to be in good agreement with previous work.  For instance,
both \citet[][]{berlindetal06b} and \citet[][]{vandenboschetal06}
suggest $\sigma_8\approx 0.8$, though no errors on this estimate are
made.  On the other hand, \citet[][]{rinesetal06} analyze a local
cluster sample selected from the SDSS and report
$\sigma_8=0.92^{+0.24}_{-0.19}$, similar to the result from
\citet[][]{bahcalletal03} and that of \citet[][]{bahcallbode03}.

The work that is perhaps most directly comparable in spirit to ours is
that of \citet[][]{gladdersetal06}, who find
$\sigma_8=0.67^{+0.18}_{-0.13}$ using a self-calibration technique to
analyze cluster abundances as observed in the Red-Sequence Survey
(RCS). As in our work, their value for $\sigma_8$ is marginalized over
uncertainties in both the mass-richness relation and the hubble
parameter $h$.  This is particularly important as it is known that
these are the parameters that are most highly degenerate with
$\sigma_8$ \citep[see e.g.][]{rozoetal04}, yet they are often held
fixed in cluster abundance studies. Despite the fact that the
central value for $\sigma_8$ in \citet[][]{gladdersetal06} is below
our $99\%$ confidence lower limit $\sigma_8>0.68$, given the large
error bars in both of our determinations it is clear that our results
are, in fact, quite consistent with each other. Moreover, the
likelihood function in \citet[][]{gladdersetal06} has a large tail
that extends out to $\sigma_8\approx1.2$, implying the overlap
between the two analysis is even larger than what the $68\%$
confidence error bars might suggest.

What is clear at this time is that a precise determination of
$\sigma_8$ still eludes us. This is unfortunate, as our large errors
for $\sigma_8$ imply that almost any reasonable value that we could find would
automatically be consistent with the $\Lambda CDM$ interpolated value
found by \citet[][]{wmap06}, $\sigma_8=0.74^{+0.05}_{-0.06}$.  It is
evident that an improvement of at least a factor of two in the error
bars is needed before $\sigma_8$ constraints from cluster abundances
become a potential probe of the evolution of dark energy.
Fortunately, such an improvement may be possible in the near future.
For instance, we are now in the process of measuring weak lensing
masses for clusters of fixed richness $\Nobs$ using the method
presented in \citet[][]{johnstonetal04} .  In principle, such a
measurement represents a direct measurement of the richness-mass
relation, and hence can place direct constraints on the mass scale
$M_1$ and the product $\alpha\beta$, the effective slope of the
mass-richness relation.  




An additional interesting result at this stage is the low value for
$\alpha$, the slope of the HOD, that we recovered with our analysis.
Whereas we find $\alpha=0.83\pm0.06$, there is a significant body of
evidence that points towards larger values of $\alpha$.  From the
observational side, \citet[][]{kochaneketal03} obtain a slope
$\alpha\approx 1$ or a little steeper on the basis of a 2MASS cluster
sample obtained from a matched filter algorithm, though we note that a
recent new analysis based on stacked X-ray emission from these
clusters suggests a lower value $\alpha=0.87\pm0.05$
\citep[][]{daietal06}.  Likewise, \citet[][]{tinkeretal06} find a
slope of unity based on the galaxy angular correlation function and
the distribution of voids in the SDSS.  Finally,
\citet[][]{zehavietal05} find that the slope $\alpha$ is close to
unity, but steadily increases with increasing luminosity.  They also
note, however, that this result is dependent on the detailed form of
the HOD parameterization.  In particular, an alternate
parameterization from \citet[][]{kravtsovetal04} with slope of unity
on the high mass end is seen to be consistent with the data as well.
From the theoretical side, \citet[][]{kravtsovetal04} showed that the
slope $\alpha$ characterizing the HOD of subhalos is very close to
unity.  Moreover, numerical simulations of galaxy formation suggest
there is an excellent correspondence between dark matter subhalos and
galaxies \citep[][]{gaoetal04,nagaikravtsov05,weinbergetal06}.  This
is further evidenced by the fact that simple models used to assign
luminosities to galaxies provide excellent matches to the luminosity
dependent galaxy-galaxy and galaxy-mass correlations functions
\citep[][]{tasitsiomietal04,mandelbaumetal05,conroyetal06,valeostriker05},
as well as the galaxy three-point function (Marin et al. 2007, in preparation).

There are, however, some lines of evidence for a lower value for
$\alpha$.  In particular, some numerical simulations suggest a value
$\alpha\approx 0.8$ on the high mass end \citep[][]{berlindetal03}.
Also, \citet[][]{scoccimarroetal01} cited a low value of the slope
$\alpha=0.8$ on the basis of the galaxy three-point function, though
we note they did not differentiate between central and satellite
galaxies, which would tend to lower the recovered slope for the HOD.
\citet[][]{abazajianetal05} also found that the best fit value for
$\alpha$ was $\alpha=0.83^{+0.22}_{-0.23}$ on the basis of a joint
analysis of the galaxy angular correlation function and the CMB,
though note the rather large error bars. Finally, analysis of the
2dFGRS cluster catalog of \citet[][]{yangetal05a} also suggests that
if $\sigma_8\approx 0.9$ then the number of galaxies in massive halos
is low relative to standard models, consistent with a lower value for
$\alpha$ \citep[][]{yangetal05b}.  Nevertheless, they emphasize their
data is equally well fit by a low $\sigma_8$ model with a standard
galaxy population, and in their most recent analysis they argue that a
moderately low $\sigma_8$ (roughly $0.8\lesssim \sigma_8 \lesssim
0.9$) is likely to be correct \citep[][]{vandenboschetal06}.

Overall, the majority of the evidence available suggests that the
slope of the HOD is typically closer to unity than what we have found.
Of course, in general one expects that in detail, the slope of the HOD will
depend on the particular criteria for galaxy selection in computing the
HOD.  At this point, about all we can say is that the maxBCG cluster sample provides
some marginal evidence for $\alpha<1$, though this is only about a $2\sigma$
effect, and, moreover, the deviation from unity might be slightly overestimated 
if indeed the selection
function in the simulations and in the real data are different.


\subsection{Additional Sources of Systematics}
\label{sec:additional_systematics}

An absolutely fundamental assumption about our statistical method is
that the cluster selection function is an inherent property of the
cluster-finding algorithm used.  More specifically, we have assumed
that given a halo of richness $\Nt$, the cluster-finding algorithm has
a probability $P(\Nobs|\Nt)$ of detecting such a halo as a cluster of
richness $\Nobs$.  If this probability is \it not \rm a property of
the cluster-finding algorithm (ie, if it is a strong function of
cosmology), then our analysis needs to be generalized, and the cosmological
dependence of the probability matrix needs to be calibrated.
Consequently, the extent to which the above probability is robust to moderate
changes in cosmology is
an inherent systematic of our method, and clearly warrants further
investigation.  We emphasize, however, that
this is true of \it any \rm characterization of a cluster selection
function obtained through the use of simulations.  That is to say, it
is not guaranteed a priori that two different yet realistic simulations will
result in the same cluster selection function. 

In addition to the above systematic, by far our most important
uncertainties are due to possible selection effects not included
within our model.  For instance, we expect each of the components of
our model --- the completeness and purity functions and the signal
matrix --- to have some redshift and richness dependence.  In this work, we have
simply ignored this possibility, though we note that due to the small
range of redshifts probed, we do not expect this systematic to be
particularly significant.  Moreover, we have explicitly demonstrated
that ignoring such evolution in the simulations does not bias our
results in any way \citep[][]{rozoetal07a}.

In addition to these selection function systematics, there are
additional sources of error which we have not included.  For instance,
in this work we have ignored possible evolution in the HOD.
Given the relatively narrow redshift range considered, and the
fact that there appears to be little
evolution in the way galaxies populate halos between redshifts $z=0$
and $z=0.8$ \citep[][]{yanetal03}, we do not expect the
no evolution assumption to be a limiting factor in our analysis.

A more theoretical systematic which we have not considered has to do
with the current uncertainty in the predicted halo mass function.  In
particular, while the halo mass function appears to be universal with
about a $\approx 20\%$ margin of error \citep[][]{jenkinsetal01},
additional work is required to test whether the halo mass function is
indeed universal to higher accuracy --- and if not, to characterize
any intrinsic cosmological dependences.  In this work, we have ignored
these complications and have made no attempt to marginalize over the
corresponding uncertainties (as is customary in the literature).  We have,
however, explicitly checked that changing the parameterization
of the halo mass function from that of \citet[][]{jenkinsetal01} to that of
\citet[][]{shethtormen02} or that of \citet[][]{warrenetal05} changes the
expectation values of our parameters by much less than our
quoted $1\sigma$ uncertainty.  

Yet another possible systematic of theoretical origin has to do with
our assumptions about the HOD.  In particular, if the HOD has any
curvature over the mass range probed, then our model will necessarily
result in biased parameter estimation.  Since at this time we are only
probing roughly one and a half decades in mass, we do not expect this to be a
significant problem.  Nevertheless, once the selection function for
the maxBCG cluster-finding algorithm is better understood, it would be
interesting to investigate to what extent the data can constrain
deviations from linearity in the mean relation between halo mass and
$\Nt$.


\subsection{Future Work and Improvements}
\label{sec:future_work}

Clearly, one of the most important problems to work on at this time is
improving our understanding of the maxBCG selection function.  Note
that this work must involve investigating a large range of
cosmological parameters so that any cosmology dependences inherent to
the cluster-finding algorithm can be adequately calibrated.  In
addition, it is possible the simulations themselves need to be refined to
produce more realistic skies so that differences in the cluster selection
function between the simulations and the real sky are minimized.

Along this same line of reasoning, an important possibility that must
be considered is to ask what the most useful richness definitions are,
and its related question, what the best way of matching halos to
clusters is.  While we have done some preliminary work in this
direction \citet[see][]{rozoetal07a}, we note that membership-based
matching algorithms obviously depend on what we mean by a halo
member.  Thus, if the definition for $\Nt$ changes, the matching of
halos to clusters, i.e. the selection function, changes. Indeed,
preliminary work suggests that uncertainties in the selection function
of the cluster-finding algorithm are significantly reduced if only
ridgeline galaxies are considered as halo members.  Note that from a
theoretical perspective, such a definition for $\Nt$ would be
perfectly fine, since galaxy formation models suggest that early type
galaxies in clusters also obey a simple Poisson HOD
\citep[][]{zhengetal05}.  Of course, the important thing is not the
fact the HOD is Poisson, but that we have a theoretically
well-motivated reason for choosing a particular form for the scatter
between halo mass and richness.

Finally, it is evident that perhaps the single most important step at this
time will be to include additional data that will soon be available for 
the maxBCG sample.  Specifically, we are now in the process of computing
ensemble-averaged X-ray, dynamical, and weak lensing masses for the 
SDSS maxBCG cluster sample.  These new data sets will directly constrain
the richness-mass relation of the clusters, and their inclusion will provide 
an entirely new tool that will further improve the cosmological constraints
derived here.

In light of the above discussion, it is clear that there remains an enormous
amount of work to be done to fully realize the potential of optical cluster 
catalogs.  Nevertheless, we emphasize that, even in this first, roughest
attempt to recover science from the SDSS maxBCG cluster sample,
we have been able to derive robust cosmological constraints that are
competitive with other approaches,  squarely placing optical 
cluster science in the general toolkit of the precision cosmology effort.

\acknowledgments 

ER would like to thank Scott Dodelson and Andrey Kravtsov for a
careful reading of the manuscript, and for many illuminating comments
that have greatly improved the quality and presentation of this work.
ER would also like to thank Wayne Hu, Zhaoming Ma, Andrew Zentner, and
Marcos Lima for useful conversations.  This work was carried out as
part of the requirements for graduation at The University of Chicago.
ER was partly
supported the Center for Cosmology and Astro-Particle Physics (CCAPP)
at The Ohio State University.  ER was also funded in part by the Kavli
Institute for Cosmological Physics (KICP) at The University of Chicago. 
RHW was primarily supported by NASA through a Hubble Fellowship
awarded by the Space Telescope Science Institute, which is operated by
the Association of Universities for Research in Astronomy, Inc, for
NASA, under contract NAS 5-26555.  RHW was also supported in part by the 
U.S. Department of Energy under
contract number DE-AC02-76SF00515.   
AEE was supported by NASA NAG5-13378, by NSF ITR grant ACI-0121671, 
and by the Miller Foundation for Basic Research in Science at UC, Berkeley.
T. McKay, A. Evrard, and B. Koester gratefully acknowledge support from 
NSF grant AST 044327.
The research described in this paper was performed in part at the
Jet Propulsion Laboratory, California Institute of Technology, under a contract
with the National Aeronautics and Space Administration.
This study has used data from the
Sloan Digital Sky Survey (SDSS, http://www.sdss.org/).  Funding for
the creation and distribution of the SDSS Archive has been provided by
the Alfred P. Sloan Foundation, the Participating Institutions, the
National Aeronautics and Space Administration, the National Science
Foundation, the U.S. Department of Energy, the Japanese
Monbukagakusho, and the Max Planck Society.  This work made extensive
use of the NASA Astrophysics Data System and of the {\tt astro-ph}
preprint archive at {\tt arXiv.org}.

\bibliographystyle{apj}
\bibliography{v8b}
\newcommand\AAA[3]{{A\& A} {\bf #1}, #2 (#3)}
\newcommand\PhysRep[3]{{Physics Reports} {\bf #1}, #2 (#3)}
\newcommand\ApJ[3]{ {ApJ} {\bf #1}, #2 (#3) }
\newcommand\PhysRevD[3]{ {Phys. Rev. D} {\bf #1}, #2 (#3) }
\newcommand\PhysRevLet[3]{ {Physics Review Letters} {\bf #1}, #2 (#3) }
\newcommand\MNRAS[3]{{MNRAS} {\bf #1}, #2 (#3)}
\newcommand\PhysLet[3]{{Physics Letters} {\bf B#1}, #2 (#3)}
\newcommand\AJ[3]{ {AJ} {\bf #1}, #2 (#3) }
\newcommand\aph{astro-ph/}
\newcommand\AREVAA[3]{{Ann. Rev. A.\& A.} {\bf #1}, #2 (#3)}

\appendix

\section{Definition of the Nuisance Parameters}
\label{app:pardef}

Here we summarize the expressions that define the nuisance parameters in
our model.  For a discussion of where these expressions come from, we
refer the reader to \citet[][]{rozoetal07a}.  The probability distribution
$P(\Nobs|\Nt)$ is characterized as follows: let $\mu_{ML}(\Nt)$ be
the most likely value for $\Nobs$ given $\Nt$.  $\mu_{ML}$ is parameterized
as a power law
\begin{equation}
\mu_{ML}(\Nt)=20\exp(B_0)(\Nt/20)^\beta,
\end{equation}
which defines $B_0$ and $\beta$.  The mean value of $\Nobs$ at fixed
$\Nt$ is found to be given by
\begin{equation}
\avg{\Nobs|\Nt} = 20 \exp(B_0+0.14)(\Nt/20)^{(\beta-0.12)}
\end{equation}
and the variance is given by 
\begin{equation}
\mbox{Var}(\Nobs|\Nt)=\exp(-3B_0+B_1)\mu(\Nt)
\end{equation}
which defines $B_1$.  The confidence regions in the $B_0-B_1$ plane for each 
of the three simulations considered in \citet[][]{rozoetal07a} are seen in figure
\ref{fig:priors}.  The solid lines define a band over which the parameters 
$B_0$ and $B_1$ are allowed to vary. 


\begin{figure}[t]
\begin{center}
\epsscale{1.2}
\plotone{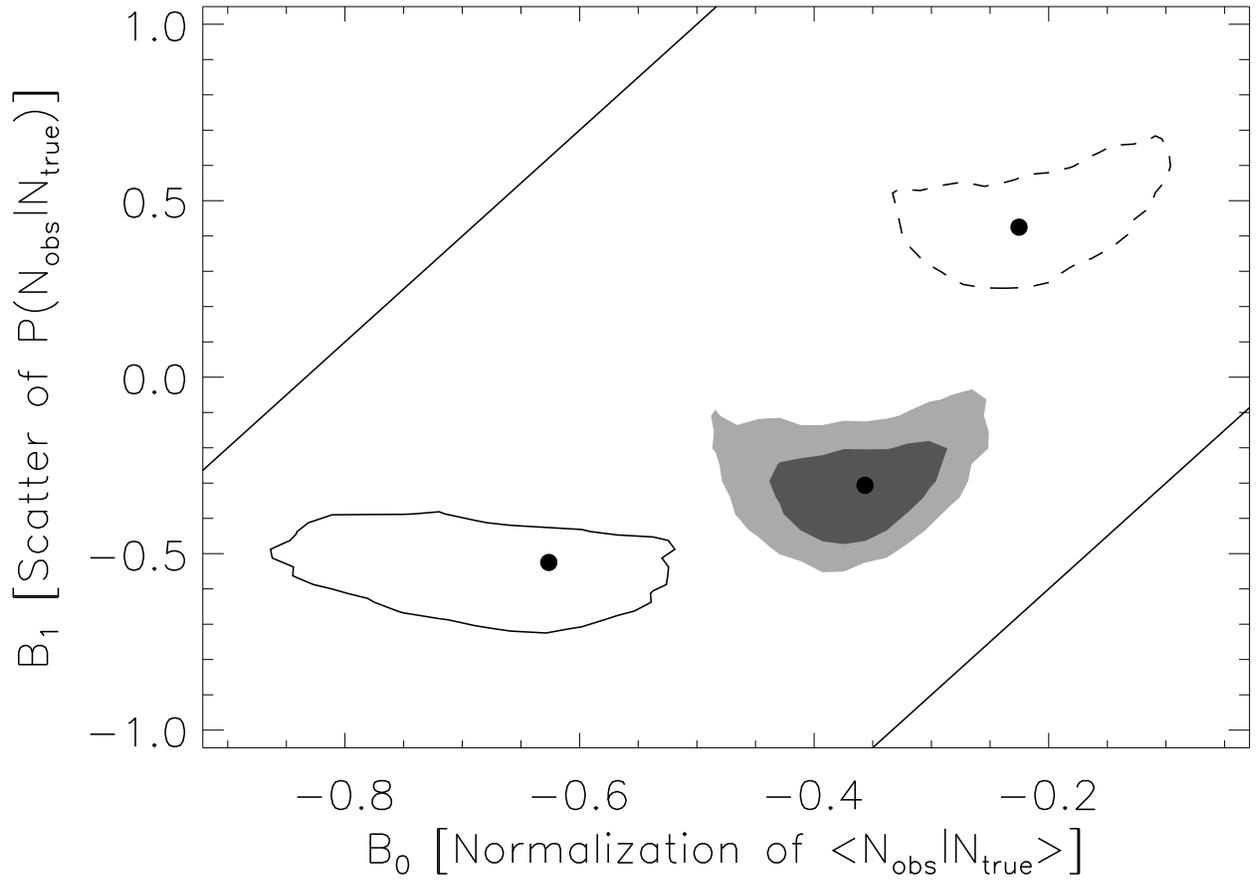}
\end{center}
\caption[Confidence Regions of the Signal Matrix Parameters]{$95\%$
  confidence regions for the $B_0$ and $B_1$ in each of the three mock 
  catalogs analyzed in \citet[][]{rozoetal07a}.  The
  dashed and solid contours are for Mocks A and B respectively.  The
  shaded contours are $68\%$ and $95\%$ confidence regions in Mock
  C. The small filled circles mark the best-fit parameters from the
  mock catalogs, and were used to generate the Monte-Carlo realizations
  from which the confidence regions are derived.  The solid lines mark
  the band over which the parameters $B_0$ and $B_1$ are allowed to
  vary. }
\label{fig:priors}
\end{figure} 


The completeness function is observed to be richness independent, and is thus parameterized
in terms of a single parameter $c$ such that $c(\Nt)=c$.  The purity function is characterized with
two parameters $p_0$ and $p_1$ such that
\begin{equation}
p(\Nobs) = \exp(-x(\Nobs)^2)
\end{equation}
where
\begin{equation}
x(\Nobs) = p_0+p_1\left(\frac{\ln(15)}{\ln(\Nobs)}-1\right).
\end{equation}

Finally, the photometric redshift distribution $\rho(z_c|z_h)$ where $z_c$
is the observed photometric redshift of a cluster and $z_h$ is the true
spectroscopic redshift of the parent halo is taken to be of the form
\begin{equation}
\rho(z_c|z_h) = \frac{1}{z_h}\rho_b(b|z_h)
\end{equation}
where $b=z_c/z_h$ is the photometric redshift bias.  $\rho_b$ is found to be Gaussian,
and the nuisance parameters $\avg{b_z}$ and $\sigma_z$ correspond to the
mean and standard deviations of said Gaussian.

\end{document}